# Progressive Generative Adversarial Binary Networks for Music Generation

Manan Oza[1], Himanshu Vaghela[1], Kriti Srivastava[1]

[1] D. J. Sanghvi College of Engineering, Department of Computer Engineering,
Mumbai, India
{manan.oza0001@gmail.com, himanshuvaghela1998@gmail.com,
kriti.srivastava@djsce.ac.in}

**Abstract.** Recent improvements in generative adversarial network (GAN) training techniques prove that progressively training a GAN drastically stabilizes the training and improves the quality of outputs produced. Adding layers after the previous ones have converged has proven to help in better overall convergence and stability of the model as well as reducing the training time by a sufficient amount. Thus we use this training technique to train the model progressively in the time and pitch domain i.e. starting from a very small time value and pitch range we gradually expand the matrix sizes until the end result is a completely trained model giving outputs having tensor sizes [4 (bar) × 96 (time steps) × 84 (pitch values) × 8 (tracks)]. As proven in previously proposed models deterministic binary neurons also help in improving the results. Thus we make use of a layer of deterministic binary neurons at the end of the generator to get binary valued outputs instead of fractional values existing between 0 and 1.

## 1  Introduction

Generating and composing music in symbolic domain using neural networks is an active research field over recent years. Efforts have been made to generate music in the form of monophonic notes [1] or lead sheets [2]. Other music generation attempts are using symbolic domain and music transcription [11, 12]. In [3] , number of instruments for the synthesized music are increased. In order to increase polyphony, music representation in the form of piano-rolls is used. Piano-rolls contains musical patterns of N-tracks in the form of pitches and corresponding time step in the form of binary matrix.

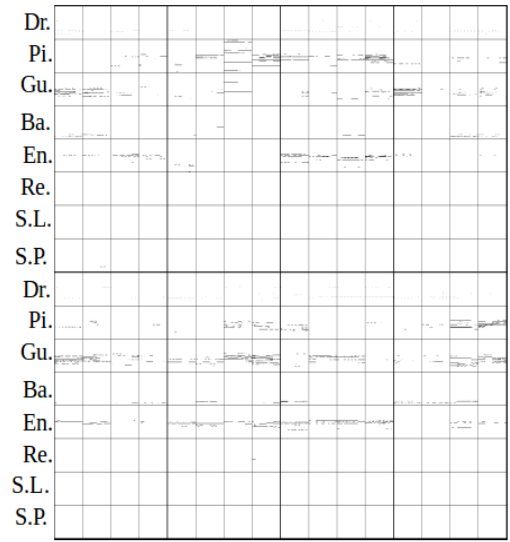

Fig. 1: Two 8 track piano-roll samples created by our proposed model.

More than one instruments make it difficult to create piano rolls due to possibility of many active notes in a time step. Recurrent neural networks and convolutional neural networks have been used recently in music generation. It is better to learn local musical patterns in CNNs whereas RNNs have been used to learn temporal feature dependency of music. In [3], five track piano-rolls were generated by using convolutional generative adversarial networks (GANs) [4]. While, the music generated was not comparable to that of human musicians, this was the first model to generate polyphonic and multi track music.

Music synthesis is further improved by using binary neurons (BN) [5, 6]. Hard thresholding (HT) and Bernoulli sampling (BS) can be applied to the floating-point predictions in BNs where BS is implemented after considering them as probabilities and HT binarizes them. While there may be many exceptions in either case. In HT, problem arises when there are many floating-point prediction values close to the threshold. In BS, binary approximation of 0 and one may be uneven.

In [2], as binarization of output of generator is done only at the testing time, discriminator D distinguishes the generated and real piano-rolls more effectively. In binary neurons, binarization is performed during training and testing both which assist the discriminator in extracting features relevant to music. The input of the discriminator is in binary form instead of floating-point. The model space $R^M$ is reduced to $2^M$ such that M is the product of the number of possible pitches and number of time steps. Relatively smaller model space makes training easy.

In [5], a refiner network R which uses binary neurons is placed between generator G and discriminator D where R binarizes the floating-point predictions made by G. Two-stage training is conducted where first, G is fixed after pretraining G and D. Second, R is trained followed by fine-tuning D. As compared to Musegan [3], this model is more effective because of the use of deterministic binary neurons (DBNs). In our proposed model, we use progressive generative adversarial networks [7] with DBNs. Our model consists a total of 12 layers in the shared generator and discriminator network and 8 layers in the refiner network at the end of all phases. Pitch and time-step values are increased progressively layer by layer. Experimental results indicate that the final output is more efficient due to the progressive training of GANs.

## 2 Background

### 2.1 Generative Adversarial Networks

Generative Adversarial Networks (GAN) have a latent space derived from the original dataset and generator tries to fool the discriminator attempting to generate realistic data as mentioned earlier. The generator function G and discriminator function D are the two main components of a GAN which are locked into a minimax game. The discriminator takes in the output of the generator or the original dataset x as input. During the training phase it learns to discern between fake and real samples. The generator takes in input as a noise vector **z** which is a sample of the prior distribution $p_z$ of the original dataset. The generator fools the discriminator with it's counterfeit sample G(**z**). The generator and discriminator are trained using a deep neural network. Wasserstein GAN (WGAN) [8] which is an alternative form of GAN measures the Wasserstein distance between the real distribution and the distribution of the generator. This distance acts as a critic to the generator function. The WGAN objective function is given as:

$$\min_G \max_D \mathbb{E}_{x p_d}[D(x)] - \mathbb{E}_{x p_z}[D(G(z))], \qquad (1)$$

here $p_d$ denotes the distribution of real data.

A new *gradient penalty* (GP) term was added by Gulrajani et al. [9] The GP term enforces the Lipschitz constraints on the discriminator and is an important factor in training of the discriminator. Thus the *gradient penalty* term added to the objective function of the GAN is:

$$E_{\hat{x} p_{\hat{x}}}\left[\left(\nabla_{\hat{x}}\|\hat{x}\| - 1\right)^2\right] \qquad (2)$$

where $p_{\hat{x}}$ is defined as sampling uniformly along straight lines between pairs of points sampled from $p_d$ and the model distribution $p_g$. It was observed that WGAN-GP [3] stabilized the training and attenuated the mode collapse issue in comparison to the weight clipping methodology used in the original WGAN. Hence we use WGAN-GP in our proposed framework.

### 2.2 Progressive Growing of GANs

In progressive growing of GANs [7] we train the GAN network in multiple phases. In phase 1, it takes in the noise vector **z** and uses $n$ convolution layers to generate a low resolution music sample. Then, we train the discriminator with the generated music and the real low resolution dataset. Once the training stabilizes, we add $n$ more convolution layers to up sampling the music to a slightly higher resolution and n more convolution layers to down sampling music in the discriminator. Here by resolution of music we imply the number of time steps and pitch values and we have taken $n = 1$. Large number of time steps and pitch values correspond to higher resolution and vice versa.

The progressive training speeds up and stabilizes the regular GAN training methods. Most of the iterations are done at lower resolutions, and training is significantly faster with com- parable music quality using other approaches. In short, it produces higher resolution images with better music quality. The progressive GAN technique uses a simplified minibatch discrimination to improve the diversity of results. Progressive GAN computes the standard deviation for each feature in each spatial location over the minibatch. Then it averages them to yield a single scalar value. It is concatenated to all spatial locations and over the minibatch at one of the latest layers in the discriminator. If the generated music samples do not have the same diversity as the real music samples, this value will be different and therefore will be penalized by the discriminator.

Progressive GAN initializes the filter weights with $\mathcal{N}(0, 1)$ and then scale the weights at runtime for each layer $\hat{w}_i = w_i / c$ where,

$$c = \left( \sqrt{\frac{2}{number\,of\,inputs}} \right)^{-1} \quad (3)$$

For the generator, the features at every convolution layers are normalized, given by:

$$b_{x,y} = \frac{a_{x,y}}{\sqrt{\frac{1}{N} \sum_{j=0}^{N-1} \left(a_{x,y}^j\right)^2 + \epsilon}} \quad (4)$$

### 2.3 Deterministic Binary Neurons

Binary Neurons (BNs) as the name suggests, give binary values as their output. In our proposed model we make use of *deterministic binary neurons* (DBNs). DBNs behave like hard thresholding function as the activation function. For a real valued input x the output of a DBN is given as:

$$DBN(x) = u(\sigma(x) - 0.5) \tag{5}$$

with $u()$ being the unit step function and $\sigma()$ being the sigmoid function.

### 2.4 Straight-Through Estimator

The function DBN given in Eq. (5) is non-differentiable and hence calculating the precise gradients for DBNs is very difficult. Thus we cannot use the conventional back propagation method to train the parameters of the network. To tackle this problem we use the sigmoid-adjusted straight through estimator proposed by Chung et al. [10] In a straight-through estimator the non-differentiable functions used in the forward pass are replaced by differentiable functions in the backward pass [11]. Functions in the backward pass are treated as identity functions and their gradient is and thus ignored. In sigmoid-adjusted straight through estimators the backward pass gradients are multiplied with the derivative of sigmoid function.

## 3 Proposed Model

### 3.1 Input Data

Just as it is used in [3, 5] we use the we use the multi track piano-roll representation which consists of music from 8 different instruments. A multi-track piano-roll is a set of piano rolls for the various representatoinal tracks (or instruments used in the music). Each piano-roll is represented as a binary-valued score-like matrix, where the horizontal and vertical axes represent time (time step) and pitch, respectively. each of the values indicate that a note is present at the particular time step. We discard the tempo information from the temporal axis and hence all beats have the same length irrespective of tempo value given.

### 3.2 Generator

In our proposed model we make use of a shared generator network $G_s$. The $G_s$ is common for all the 8 tracks and is progressive as shown in Fig. 2 . The progressive network starts with the first two layers from Fig. 5. After every two epochs one layer is added to the model and the variables of the previous iterations are loaded into the layers carried forward from the previous model. Variables for the new layers are randomly initialized and normalized using the formula in Eq. 4. The first layer is a fully connected layer where as the second layer reshapes the tensor passed on from the first layer. All subsequent layers in the generator are transposed 3D convolution layers.

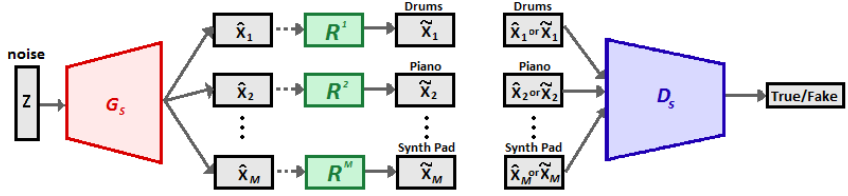

Fig. 2 Our proposed architecture. The shared generator network $G_s$ and shared discriminator network $D_s$ are trained progressively with a refiner network having variable tensor size in between.

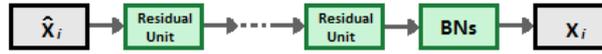

Fig. 3: The refiner network with the tensor size same throughout the network and deterministic binary neurons.

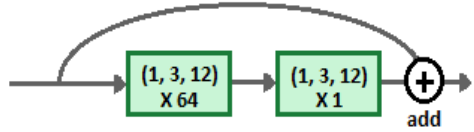

Fig. 4: Residual unit used in the refiner network. It's values denote the kernel size and the number of the output channels of the two convolutional layers.

### 3.3 Refiner

The refiner network R is the second part of the generator network and is composed of M private networks each denoted by $R_i$, i = 1, ..., M, one for each track (instrument). The refiner network takes in input from the generator and split it along the track axis. The task of the refiner is to refine the real-valued outputs from the generator, $\hat{x} = G(z)$ into binary values denoted by $\tilde{x}$. Thus the refiner network does not learn any new feature or mapping from the generator onto the dataset. The refiner output consists of the output merged from all the tracks along the track axis. The refiner consists of 2 residual blocks followed by a layer of DBNs.

```
------------------------------------ < GAN > -----------------------------------
------------------------------------ GAN/G -------------------------------------
Input                                              (32, 4, 96, 84, 8)
--------------------------------------------------------------------------------
[GAN/G/shared]
Input                                              (32, 128)
GAN/G/shared/Layer_0              dense            (32, 32)
GAN/G/shared/Layer_1              reshape          (32, 4, 1, 1, 8)
GAN/G/shared/Layer_2              transconv3d      (32, 4, 3, 1, 8)
GAN/G/shared/Layer_3              transconv3d      (32, 4, 6, 1, 8)
GAN/G/shared/Layer_4              transconv3d      (32, 4, 12, 1, 8)
GAN/G/shared/Layer_5              transconv3d      (32, 4, 24, 1, 8)
GAN/G/shared/Layer_6              transconv3d      (32, 4, 48, 1, 8)
GAN/G/shared/Layer_7              transconv3d      (32, 4, 96, 1, 8)
GAN/G/shared/Layer_8              transconv3d      (32, 4, 96, 2, 8)
GAN/G/shared/Layer_9              transconv3d      (32, 4, 96, 4, 8)
GAN/G/shared/Layer_10             transconv3d      (32, 4, 96, 12, 8)
GAN/G/shared/Layer_11             transconv3d      (32, 4, 96, 84, 8)
--------------------------------------------------------------------------------
[GAN/G/refiner0]
Input                                              (32, 4, 96, 84, 1)
GAN/G/refiner0/Layer_0            identity         (32, 4, 96, 84, 1)
GAN/G/refiner0/Layer_1            identity         (32, 4, 96, 84, 1)
GAN/G/refiner0/Layer_2            conv3d           (32, 4, 96, 84, 64)
GAN/G/refiner0/Layer_3            conv3d           (32, 4, 96, 84, 1)
GAN/G/refiner0/Layer_4            identity         (32, 4, 96, 84, 1)
GAN/G/refiner0/Layer_5            identity         (32, 4, 96, 84, 1)
GAN/G/refiner0/Layer_6            conv3d           (32, 4, 96, 84, 64)
GAN/G/refiner0/Layer_7            conv3d           (32, 4, 96, 84, 1)
GAN/G/refiner0/Layer_8            identity         (32, 4, 96, 84, 1)
--------------------------------- GAN/D ----------------------------------------
Input                                              (32, 4, 96, 84, 8)
--------------------------------------------------------------------------------
[GAN/D/shared]
Input                                              (32, 4, 96, 84, 8)
GAN/D/shared/Layer_0              conv3d           (32, 4, 96, 12, 8)
GAN/D/shared/Layer_1              conv3d           (32, 4, 96, 4, 8)
GAN/D/shared/Layer_2              conv3d           (32, 4, 96, 2, 8)
GAN/D/shared/Layer_3              conv3d           (32, 4, 96, 1, 8)
GAN/D/shared/Layer_4              conv3d           (32, 4, 48, 1, 8)
GAN/D/shared/Layer_5              conv3d           (32, 4, 24, 1, 8)
GAN/D/shared/Layer_6              conv3d           (32, 4, 12, 1, 8)
GAN/D/shared/Layer_7              conv3d           (32, 4, 6, 1, 8)
GAN/D/shared/Layer_8              conv3d           (32, 4, 3, 1, 8)
GAN/D/shared/Layer_9              conv3d           (32, 4, 1, 1, 8)
GAN/D/shared/Layer_10             reshape          (32, 32)
GAN/D/shared/Layer_11             dense            (32, 1)
--------------------------------------------------------------------------------
```

Fig. 5: The final structure of our 12 layer proposed model with each layer, its type and output tensor size

### 3.4 Discriminator

The discriminator is the mirror replica of the generator for every step of training. Denoted by $D_s$, the discriminator takes in outputs generated by generator or the original dataset as inputs. It is made up of 3D convolution layers with reshaping done at the second last layer and the last layer having only one output i.e. true or false. When the original data is fed into the discriminator, it trains itself to distinguish between real and fake data and when the output from generator is fed it discerns whether the sample provided is real or fake (true or false). Contrary to the model proposed in [5] we do not split the discriminator into different streams because it is empirically proven that the progressive training of the generator and discriminator is sufficient enough to extract relevant the features pertaining to the tracks individually and also the music sample as a whole.

## 4   Analysis

### 4.1   Implementation and Training

We use the cleansed subset of the Lakh Pianoroll Dataset (LPD-*cleasned*) used in [5]. It consists of 2,291 multitrack (8 track to be exact) pianorolls which are 4/4 in time. Out of each song six four-bar phrases are selected which amounts to a final dataset of 13,746 phrase samples. Similar to the proposed model used in [5] we set the temporal resolution to 24 time steps per beat which covers the common temporal patterns like triplets and 32th notes. The note pitch has 84 possible values, from C1 to B7.

The eight tracks used in the LPD-*cleasned* dataset correspond to the following instruments: *Drums, Piano, Guitar, Bass, Ensemble, Reed, Synth Lead and Synth Pad*. Thus the size of the input tensor for the discriminator and the output tensor for the generator at the last phase of training (when all layers are added) is 4 (bar) × 96 (time step) × 84 (pitch) × 8 (track).

The model as created in the last phase of training is shown in Fig. 5. The input random vector has a length of 128. The optimization that we use is Adam optimization and we normalize the generator and refiner. We implement slope annealing [10] to networks with binary neurons, where the slope of the sigmoid function in the sigmoid-adjusted straight-through estimator is multiplied by 1.1 at the end of each epoch. We use a batch size of 32 thus having 429 iterations per epoch. Every phase is trained for 2 epochs i.e. after the addition of each layer the network is trained for two epochs.

As already proven in [5] DBNs produce better results than Stochastic Binary Neurons (SBNs) we make use of DBNs in our framework. We use a single training methodology with one phase starting after the addition of each layer.

### 4.2   Results

We compare our results with the previously stated Bmusegan Model proposed by Dong et al. [5]. We trained both the models for a total of approximately sixteen epochs.

The Fig. 6 and Fig. 7 show the log of absolute generator loss values for every iteration of the model proposed in [5] and the model proposed by us respectively. Our proposed model evidently shows better stability in 100 consecutive loss values as compared to the model proposed in [5] which displays large variations in the loss values. In Figure 6 the values before the black vertical line pertain to the pretraining and those after the line pertain to the training phase.

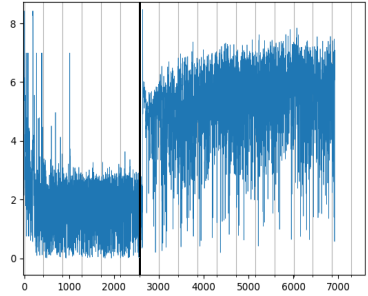
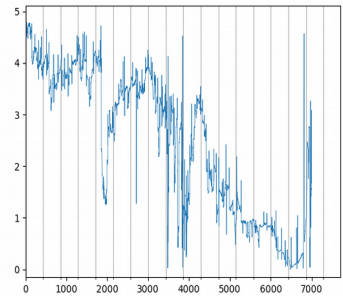

Fig. 6: Generator loss of Bmusegan model

Fig.7:Generator loss of our proposed model

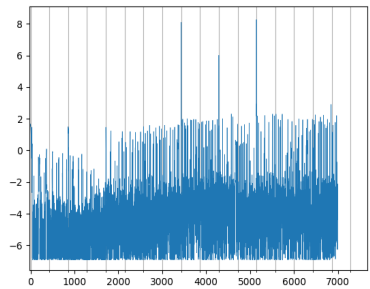
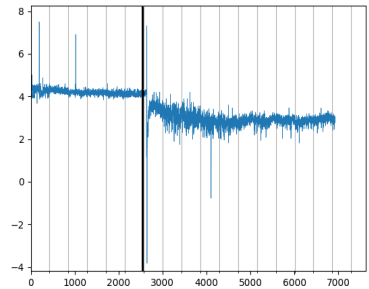

Fig. 9:Discriminator loss of our proposed model.

Fig. 8:Discriminator loss of Bmusegan model.

The vertical grid lines in the graphs denote the end of one epoch.

Fig. 7 and Fig. 8 show the log of absolute discriminator loss values for every iteration of the model proposed in [5] and the model proposed by us respectively. Our proposed model evidently shows better accuracy in distinguishing between fake and real music samples. In Fig. 9 the values before the black vertical line pertain to the pre-training and those after the line pertain to the training phase.

### 4.3  User Study

Eventually, a user study was conducted where 50 participants were randomly selected across the internet where they were asked to compare the music samples generated by our model and bmusegan [5] and vote was taken considering four parameters. Results obtained are shown in table I and II. It indicates that results in our model are efficient as compared to bmusegan.

Table 1. User study conducted for 50 participants.

|  | Bmusegan | Our model |
|---|---|---|
| pleasntness | 0.15 | 0.85 |
| harmonicity | 0.59 | 0.41 |
| rhythmicity | 0.48 | 0.52 |
| Overall rating | 0.11 | 0.89 |

Table 2. Model comparisions.

|  | Bmusegan | Our model |
|---|---|---|
| Training time for 16 epochs (on a Tesla K-80 GPU)(in hours) | 28.6 | 25.1 |
| Overall accuracy in diffrentiating between fake and real samples (in %) | 85.632 | 89.728 |

## 5 Conclusion

Our presented model for progressively training a convolutional GAN for generating binary-valued piano-rolls by using deterministic binary neurons at the output layer of the generator. We have trained our model on the eight-track Lakh Piano-roll dataset. Analysis showed that the generated results of our progressively trained model with deterministic binary neurons exhibits lesser fragmentation of the notes and improved periodicity and melodic perception. The current model does not produce music equivalent to that composed by professionals but it has definitely proven to be an improvement over the previous models. This model can further extended to generate realistic speech to text voices with life like voice modulations and in sound and digital signal refinement.